\documentclass{ws-procs9x6}
\usepackage[latin1]{inputenc}
\usepackage[T1]{fontenc}
\usepackage{graphicx}
\usepackage[dvips]{epsfig}
\usepackage{color}
\usepackage{bbm}

\definecolor{holger}{rgb}{0.7,0,0.2}

\definecolor{red}{rgb}{1,0,0}

\definecolor{alexej}{rgb}{0.8,0.1,0}

\newcommand{\const}{c}

\newcommand{\pt}{\mathcal{T}}
\newcommand{\Eqref}[1]{Eq.~\eqref{#1}}
\newcommand{\bgam}{\boldsymbol{\gamma}}

\begin{document}

\title{Geometry-Temperature Interplay in the Casimir Effect }
\author{H.~GIES}
\address{Theoretisch-Physikalisches Institut, Universit\"at Jena, D-07743
  Jena, Germany\\ 
and Helmholtz Institute Jena, D-07743 Jena, Germany}
\author{A.~WEBER}
\address{Institute for Theoretical Physics, Heidelberg University, D-69120
  Heidelberg, Germany}
\date{\today}

\begin{abstract}
We discuss Casimir phenomena which are dominated by long-range fluctuations. A prime example is given by ``geothermal'' Casimir phenomena where thermal fluctuations in open Casimir geometries can induce significantly enhanced thermal corrections. We illustrate the underlying mechanism with the aid of the inclined-plates configuration, giving rise to enhanced power-law temperature dependences compared to the parallel-plates case. In limiting cases, we find numerical evidence even for fractional power laws induced by long-range fluctuations. We demonstrate that thermal energy densities for open geometries are typically distributed over length scales of $1/T$. As an important consequence, approximation methods for thermal corrections based on local energy-density estimates such as the proximity-force approximation are expected to become unreliable even at small surface separations.

\end{abstract}

\keywords{Casimir effect, finite-temperature field theory, worldline approach}

\bodymatter

\section{Introduction}
Beyond its many attributes and genuine properties, the Casimir effect
\cite{Casimir:dh,Bordag:2001qi,Milton:2001yy} is also a phenomenon that can be dominated by long-range
fluctuations. At~first sight, this statement may seem surprising as many
standard Casimir examples do not show manifest signatures of a long-range
fluctuation phenomenon (such as, e.g., critical phenomena). For instance, the
length scale of fluctuations associated with the classic Casimir effect
between parallel plates is certainly set by the plate
separation $a$, serving effectively as a long-range cutoff.

Another seeming counter-example for the above-given statement is the
sphere-plate configuration at small separation distances $a$, which is
experimentally highly relevant \cite{Lamoreaux:1996wh}. Here, the Casimir
interaction energy behaves as $E\sim R/a^2$ for $a/R\to 0$, where $R$ is the
sphere radius. It thus exhibits a simple power law with integer coefficients
that follow from a geometric scaling analysis known as the proximity force
approximation (PFA) \cite{pft1}. (The~validity of this approximation in the
asymptotic regime has been confirmed by analytical as well as numerical
methods of quantum field theory
\cite{semicl,Gies:2003cv,Scardicchio:2004fy,Gies:2006bt,Bordag:2006vc}, see also Ref.~\refcite{Bulgac:2005ku} for a solution at larger distances.) Similar geometric
analyses work equally well for, say, electrostatic forces in the same
configuration which are not related to any long-range fluctuation phenomenon.

In the present work, we argue that these examples do not reveal the long-range
nature of the Casimir effect, as the corresponding interaction energies are
dominated by rather localized energy densities. E.g., for the sphere-plate
case, the dominant contribution to the Casimir force arises from the region
between the surfaces at closest separation. By contrast, we present Casimir
phenomena in the following where the energy density is distributed over a wide
range of scales, such that the potential long-range nature of the Casimir
effect becomes most prominent.

An example of this class of phenomena is the nontrivial interplay of
finite-temperature and geometry dependences of the Casimir effect. As first
conjectured in Ref.~\refcite{Scardicchio:2005di}, the thermal modifications of the
Casimir effect can differ qualitatively for different geometries. This is
because the thermal corrections arise from thermal excitations of the
fluctuation spectrum, which in turn depends strongly on the geometry. First
analytical as well as numerical evidence of this ``geothermal'' interplay has
been provided in Ref.~\refcite{Gies:2008zz} by applying the worldline formalism to a
perpendicular-plates configuration. A detailed study of this phenomenon for
the more general inclined-plates case has been performed in
Ref.~\refcite{Weber:2009dp}, the results of which will be used as a quantitative example for our arguments in the following.

The purpose of the present contribution is to develop the general physics picture underlying the geothermal Casimir phenomena.
In Sect.~\ref{sec:geothermal}, we discuss the origin and various general
perspectives on the interplay between geometry and finite temperature. Section
\ref{sec:worldline} briefly summarizes the worldline method which is a
powerful tool to analyze this interplay quantitatively. Several examples will
be discussed in Sect.~\ref{sec:examples}, where we also present new results for the thermal force density of specific open geometries. Conclusions are given in
Sect.~\ref{sec:conclusions}.

\section{Origin of geothermal Casimir phenomena}
\label{sec:geothermal}

The origin of a nontrivial interplay between geometry and temperature in the
Casimir effect can be understood in simple terms. Consider the
classical parallel-plate case: as the wavelengths of the fluctuations
orthogonal to the plates have to be commensurate with the distance $a$ between
the plates, this corresponding relevant part of the spectrum $\omega_n^2(a)=
p_\|^2+(\pi n/a)^2$ has a gap of wave number $k_{\text{gap}}=\pi/a$. As is
obvious, e.g., from the partition function $Z=\sum_n \exp(-\omega_n(a)/T)$,
the gapped modes are exponentially suppressed at small temperatures $T\ll
k_{\text{gap}}$. In $D=4$ spacetime dimensions, the integration over the
parallel modes $p_\|$ converts this exponential dependence into the
low-temperature power law $T^4$ for the parallel-plate Casimir force. The
corresponding thermal contribution to the interaction energy (apart from a
distance-independent term) is
\begin{equation}
\Delta E_\|(T) = \frac{\pi^2}{90}\, A\, a\, T^4, \quad aT\ll 1, \label{eq:ppthermal}
\end{equation}
where $A$ denotes the plate's area.  The above-given argument for a
suppression of thermal contributions applies to all geometries with a gap in
the relevant part of the spectrum (e.g. concentric cylinders or spheres,
Casimir pistons, etc.). These geometries are called {\em closed}.

By contrast, {\em open} geometries with a gapless relevant part of the
spectrum have no such suppression of thermal contributions. Any small value of
the temperature can always excite the low-lying modes in the
spectrum. Therefore, we expect a generically stronger thermal contribution
$\sim (aT)^\alpha$ with $0<\alpha<4$.

Another argument for the fundamental difference between open and closed
geometries and thermal corrections is the following: Eq.~\eqref{eq:ppthermal}
can also be written as $\Delta E_\|(T)= V\, \epsilon_{\text{SB}}$, where $V=aA$ is
the volume between the parallel plates, and $\epsilon_{\text{SB}}$ is the
Stefan-Boltzmann free energy density of the radiation field. Hence, we can
understand the low-temperature correction in the parallel-plate case as an
excluded volume effect: the thermal modes of the radiation field at low
temperatures do not fit in between the plates, and therefore the corresponding
volume does not contribute to the total thermal free energy. By contrast, open
geometries by construction cannot be associated with any (unambiguously
defined) excluded volume, such that significant deviations from a $T^4$
behavior can be expected.

These considerations immediately point to the possibility that the thermal part of the
low-temperature Casimir effect can be dominated by long-range
fluctuations. This is because a temperature much lower than the inverse
distance, $aT\ll 1$, sets a new length scale which can be much larger as the
plate distance as well as any other length scale of the geometry (such as a
sphere radius). In closed geometries, this length scale is effectively cut off
by the gap in the spectrum, implying the parametric suppression of thermal
effects. In open geometries, this length scale sets a relevant scale that can,
for instance, reflect the spatial extent of the distribution of the thermal
energy density. The total thermal energy thus can receive dominant
contributions from long-range modes corresponding to significantly extended
thermal energy distributions.

An important consequence can already be anticipated at this point:
approximation methods that are based on local considerations will generically
fail to predict the correct low-temperature correction in open geometries. An
example is given by the PFA which is based on the assumption that the Casimir
energy can be estimated by integrating over local parallel-plates energy
densities. Whereas this approximation may or may not work at zero temperature
depending on the geometric details of the configuration, it is even
conceptually questionable at finite temperature, as open geometries should not
be approximated by closed-geometry building blocks. Quantitatively, such a
procedure is expected to fail, as local energy-density approximations will not
be able to capture the contributions from larger length scales induced by
long-range modes.

The temperature-geometry interplay is not an academic problem:
experimentally important configurations such as the sphere-plate or the
cylinder-plate geometry belong to this class of open geometries, but thermal
corrections have so far been approximated by the PFA. Whether or not a
potentially significant geothermal interplay may exist in the relevant
parameter range $aT \sim 0.01\dots0.1$ is a technically challenging
quantitative problem.

The considerations so far have concentrated on the low-temperature limit. In
fact, the high-temperature limit exhibits a universal linear dependence on the
temperature for a simple reason. At high temperature in the imaginary-time
formalism, only the zeroth Matsubara mode can contribute as all higher modes
acquire thermal masses $\sim \pi T$ and hence are largely suppressed. The
zeroth Matsubara mode has no temperature dependence at all, such that the only
temperature dependence arises from the measure of the fluctuation trace which
is linear in $T$. A less technical argument with the same result can be based
on the underlying Bose-Einstein distribution governing the bosonic thermal
fluctuations of the radiation field. This~distribution increases as $\sim {T}$
in the high-temperature limit, inducing this linear temperature dependence directly
in the thermal energy. The properties of the geometry only enter the prefactor
in the high-temperature limit. Universal features of thermal Casimir energies
with an emphasis on the high-temperature limit have been systematically
studied in Refs.~\refcite{Balian:1977qr,Klich:2000qm,Nesterenko:2000cf,Bordag:2001jc}.

\section{Worldline method for the Casimir effect}
\label{sec:worldline}

Let us briefly review the worldline approach\cite{worldline} to the Casimir effect as it is
needed for the present line of argument. For the remainder of the paper, we
consider a fluctuating massless scalar field satisfying Dirichlet boundary
conditions on the Casimir surfaces. For details of the formalism, see
Refs.~\refcite{Gies:2003cv,Gies:2006cq,Weber:2009dp}.   Consider a configuration $\Sigma$ consisting
of two static surfaces $\Sigma_1$ and $\Sigma_2$. The worldline
representation of the Casimir interaction energy reads
\begin{align}\label{Int-1}
E_\mathrm{c}=-\frac{1}{2 (4\pi)^{D/2}} \!\int_{0}^\infty
\!\!\! \frac{\mathrm{d}\pt}{\pt^{1+D/2}} \!\int \!\mathrm{d}^d
x_\mathrm{CM}\left\langle 
\Theta_\Sigma [\mathbf{x}(\tau)] \right\rangle,
\end{align}
where $D=d+1$ denotes the spacetime dimensions.
The worldline functional obeys $\Theta_\Sigma[\mathbf{x}]=1$
if a worldline $\mathbf{x}(\tau)$ intersects both surfaces
$\Sigma=\Sigma_1\cup\Sigma_2$, and is zero otherwise.

The expectation value in \Eqref{Int-1} is taken with respect to an ensemble of
$d$-dimensional closed  Gau\ss ian worldlines with center of mass
$\mathbf{x}_{\mathrm{CM}}$,
\begin{equation}
\langle \dots \rangle = \frac
{  \int_{\mathbf{x}_{\text{CM}}} \mathcal{D} \mathbf x\,\dots\,  e^{-\frac{1}{4}
  \int_0^{\pt} \mathrm{d}\tau \, \dot{\mathbf x}^2(\tau)}}
{
  \int_{\mathbf{x}_{\text{CM}}} \mathcal{D} \mathbf x \, e^{-\frac{1}{4}
  \int_0^{\pt} \mathrm{d}\tau \, \dot{\mathbf x}^2(\tau)}}.\label{eq:expval}
\end{equation}
Equation (\ref{Int-1}) expresses the fact that all worldlines intersecting
both surfaces do not satisfy Dirichlet boundary conditions on both
surfaces. They are removed from the ensemble of allowed fluctuations by the
$\Theta$ functional and thus contribute to the negative Casimir interaction
energy. The auxiliary propertime parameter $\pt$ scales the extent of a
worldline by a factor of $\sqrt{\pt}$. Large $\pt$ correspond to long-range,
small $\pt$ to short-range fluctuations.

Finite temperature $T=1/\beta$ in the Matsubara formalism {is equivalent} to a
compactified {Euclidean} time on the interval $[0,\beta]$. The worldlines now
live on $S^1\times \mathbbm{R}^d$ and can carry a winding number. Summing over
all winding numbers, the Casimir energy (\ref{Int-1}) becomes
\begin{eqnarray}\label{Int-3}
E_\mathrm{c}=-\frac{1}{2 (4\pi)^{D/2}}  \int_{0}^\infty\!\!\!
\frac{\mathrm{d}\pt}{\pt^{1+D/2}}\!\!\left(1+2\sum_{n=1}^{\infty}\!e^{-\frac{n^2\beta^2}{4
\pt}}\right)\!\! \int \! \mathrm{d}^d x_\mathrm{CM}\left\langle
\Theta_\Sigma [\mathbf{x}(\tau)] \right\rangle.
\end{eqnarray}
The finite-temperature worldline formalism for static configurations thus
boils down to an additional winding-number prefactor in front of the worldline
expectation value. Notice that the winding-number sum is directly related to
the standard Matsubara sum by a Poisson resummation.

Finally, it is advantageous to rescale the worldlines such that the velocity
distribution becomes independent of $\pt$,
\begin{equation}
\bgam(t):= \frac{1}{\sqrt{\pt}} \mathbf x(\pt t)\quad \!\!\!\to\!\!\!\quad
e^{-\frac{1}{4} \int_0^{\pt}   \dot{\mathbf{x}}^2 \mathrm{d}\tau} = e^{-\frac{1}{4}
  \int_0^1  \dot{\bgam}^2 \mathrm{d}t},
\end{equation}
where $\dot{\bgam}=\mathrm{d}\bgam(t)/\mathrm{d}t$.  The worldline integrals
can be evaluated also numerically by Monte Carlo methods in a straightforward
manner. Various efficient ab initio algorithms for generating discretized
worldlines with Gau\ss ian velocity distribution have been developed, see,
e.g., Refs.~\refcite{Gies:2003cv,Gies:2005sb}.

\section{Geothermal phenomena for inclined Casimir plates}
\label{sec:examples}

\subsection{Inclined plates at zero temperature $T=0$}
\begin{figure}[t]
\begin{center}
\begin{minipage}[c]{5.5 cm}\includegraphics[width=1.1\linewidth]{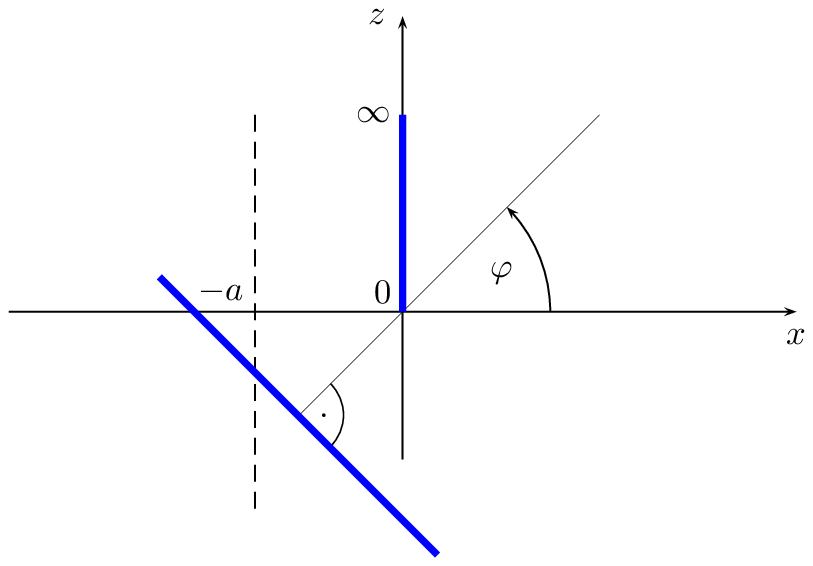}
\end{minipage} \ \ \ \
\begin{minipage}[c]{4.5 cm}
\includegraphics[width=1.1\linewidth]{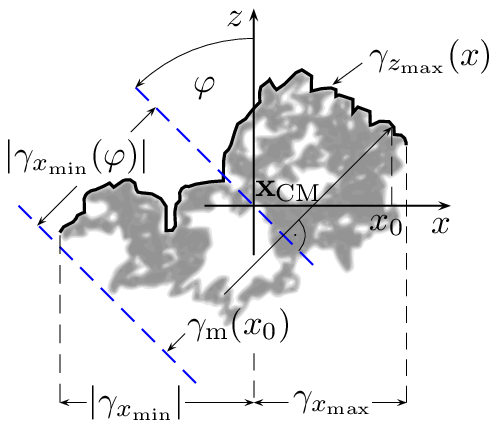}
\end{minipage}
\end{center}
\caption{Left panel: sketch of the inclined-plates configuration. The infinite
  plate (dashed line) is rotated in the $x,z$ plane by an angle $\varphi$. As
  special cases, $\varphi=0$ corresponds to the configuration of one
  semi-infinite plate parallel to an infinite plate (1si configuration),
  whereas $\varphi=\pi/2$ yields the perpendicular-plates configuration. Right
  panel: all relevant information for the evaluation of the Casimir energy
  (\ref{T0-ip-6}) of inclined plates is encoded in the function
  $\gamma_\mathrm{m}(x)$, which has to be integrated from
  $\gamma_{x_\mathrm{min}}$ to $\gamma_{x_\mathrm{max}}$.}\label{T0-ip-iloop}
\end{figure}

The inclined-plates (i.p.) configuration turns out to be an ideal work horse
to study geothermal phenomena, in particular, the transition from open to
closed geometries and the role of long-range fluctuations. It was studied in detail in
Ref.~\refcite{Weber:2009dp} for general $D$. Very recently, results for inclined plates have been obtained for the electromagnetic case at zero temperature using scattering theory\cite{Graham:2009zb}. Here, we summarize our results for $D=4$
and provide more details on the geometry-temperature interplay.

\begin{figure}[t]
\begin{center}
\begin{minipage}[c]{4.5 cm}
\includegraphics[width=1.1\linewidth]{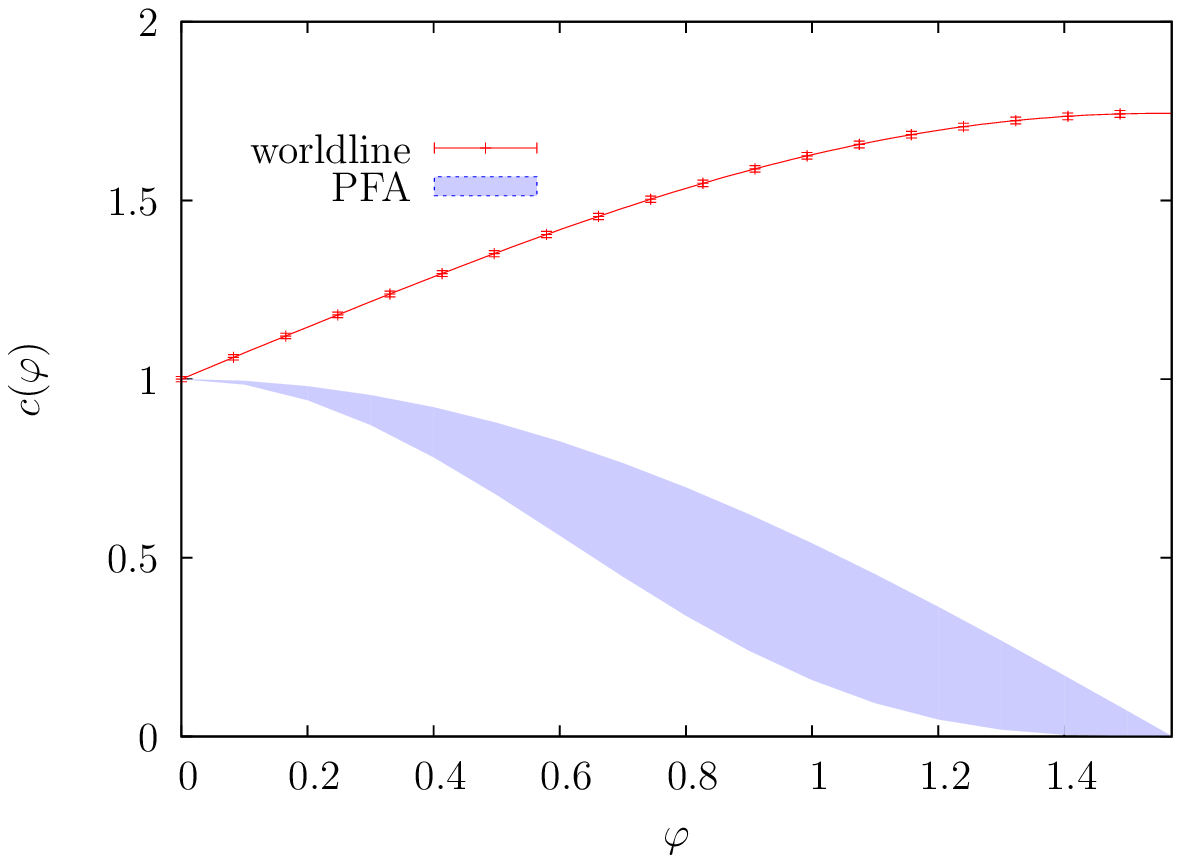}
\end{minipage} \ \ \
\begin{minipage}[c]{5.5 cm}
\includegraphics[width=1.1\linewidth]{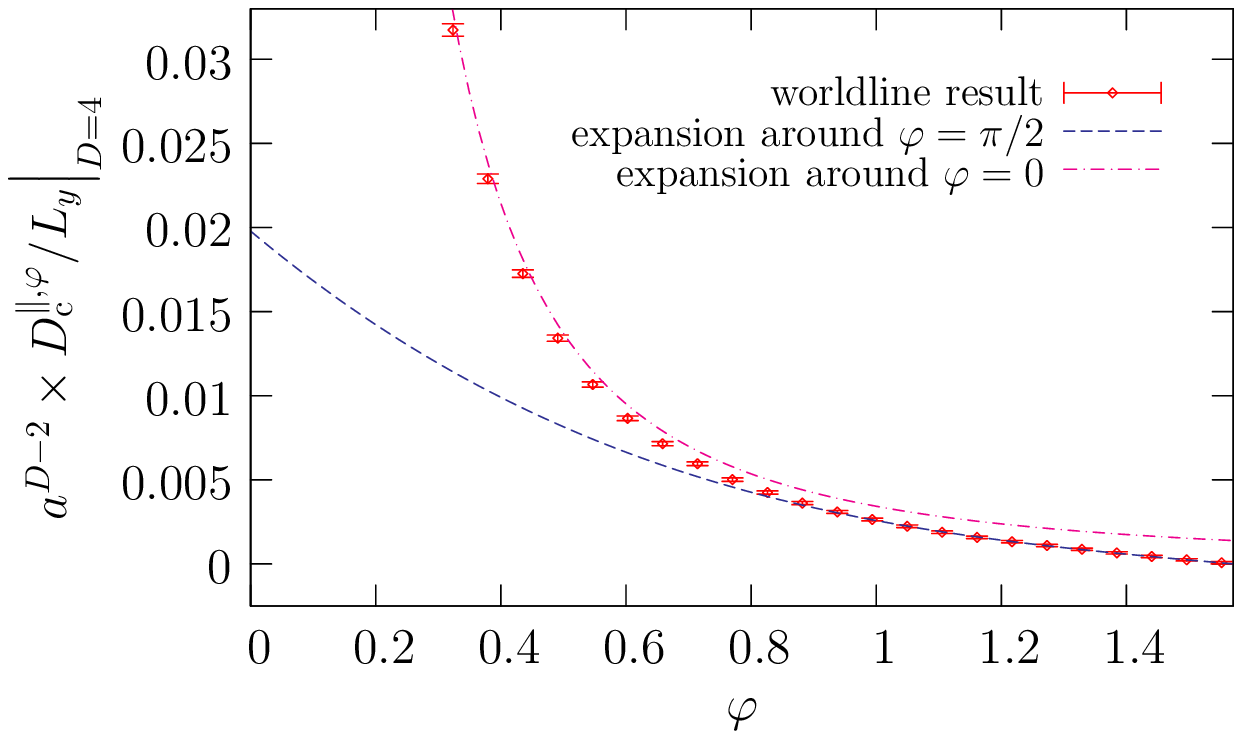}
\end{minipage}
\end{center}
\caption{Left panel: normalized Casimir energy per edge length of
  the inclined-plates configuration, $c(\varphi)=-
    2880 E_\mathrm{c}^\mathrm{i.p.,\varphi}a^{2}\sin(\varphi)/\pi^2 L_{y} $. 
%
The blue area corresponds to the PFA prediction: for the upper bound, we have used
the infinite plate as a basis, for the lower bound the semi infinite plate,
yielding $\cos(\varphi)$ and $\cos^3(\varphi)$ as bounding curves,
respectively. Right panel: normalized Casimir torque per edge length
(Eq. \ref{T0-ip-torque-1})
and its expansion around
  $\varphi=\pi/2$ {(Eq. (\ref{T0-ip-torque-2}))} and $\varphi=0$
  {(Eq. (\ref{T0-ip-torque-3}))}, respectively.
} \label{T0-ip-7}
\end{figure}

The inclined-plates configuration consists of a perfectly thin semi-infinite
plate above an infinite plate at an angle $\varphi$, see
Fig. \ref{T0-ip-iloop}. The semi-infinite plate has an edge of length
$L_{y}$. The area of the infinite plate is~$A$. The limit $L_y,\, A\to\infty$
is implicitly understood. Let $a$ be the minimal distance between the plates.
Evaluating the $\Theta$ functional for this configuration \cite{Weber:2009dp} in
(\ref{Int-1}),  the Casimir energy can be written as
\begin{align}
  \frac{E_\mathrm{c}^\mathrm{i.p.,\varphi}}{L_{y}}=&-\frac{\csc (\varphi )}{
    96\,\pi^{2}\,  a^{2}}\left\langle
  \int_{\gamma_{x_\mathrm{min}}}^{\gamma_{x_\mathrm{max}}} \mathrm{d} x \
\gamma_\mathrm{m}^{3}(x)\right\rangle,
\label{T0-ip-6}
\end{align}
where
\begin{align}
\gamma_\mathrm{m}(x)&\equiv x \cos (\varphi )+\sin
(\varphi ) \gamma_{z_{\mathrm{max}}}(x)-\gamma_{x_{\mathrm{min}}}(\varphi), \label{T0-ip-5b}
\\
\gamma_{{x_{\mathrm{min}}}}(\varphi)&
  \equiv\min_t\left(\gamma_{{x}}(t)\cos(\varphi)
  +\gamma_{z}(t)\sin(\varphi)\right). \label{T0-ip-3}
\end{align}
Here, the Casimir energy has been related to simple geometrical properties of
the worldlines: $\gamma_{{x_{\mathrm{min}}}}(\varphi)$ measures the
extent of the worldline in the negative $x$ direction of a coordinate system rotated by
the angle $\varphi$,
and $\gamma_{z_{\text{max}}}(x)$ denotes the $x$-dependent
  envelope of the worldline in positive $z$ direction,
see Fig.~\ref{T0-ip-iloop}.
Equation (\ref{T0-ip-6}) is shown as a function of $\varphi$ in
Fig. \ref{T0-ip-7}.

For $\varphi= \pi/2$ 
we rediscover
the perpendicular plates result \cite{Gies:2006xe,Klingmuller:2007} as a special case. For $\varphi=0$,
the integral in \Eqref{T0-ip-6} can be done analytically
resulting in $\langle\lambda^4/4\rangle=\pi^4/30$.  Together
with the $\varphi$-dependent prefactor, \Eqref{T0-ip-6} diverges as
$\varphi\rightarrow 0$ as it should. This is because \Eqref{T0-ip-6}
corresponds to the energy per unit \textit{edge length}, whereas for
$\varphi\to0$ the Casimir energy becomes proportional to the {\em area} of the
semi-infinite plate.
%
%
%
%

The result for the parallel limit $\varphi\rightarrow 0$ has to
arise from the general inclined-plates
formula \Eqref{T0-ip-6}, but involves a subtle limiting
process which was performed in Ref.~\refcite{Weber:2009dp}.
For this 1si case, the total Casimir interaction energy decomposes into
\cite{Gies:2006xe, Klingmuller:2007},
\begin{align}\label{T0-ip-4}
  E_\mathrm{c}^\mathrm{1si}
  =E_\mathrm{c}^\mathrm{1si,\parallel}+E_\mathrm{c}^\mathrm{1si,edge},
\end{align}
where $E_\mathrm{c}^\mathrm{1si,\parallel}/A$ is the usual Casimir energy per
unit area of two parallel plates, with $A$ now being the area
of the semi-infinite plate. The subleading edge energy
$E_\mathrm{c}^\mathrm{1si,edge}$ arises solely
due to the presence of the edge and is proportional to the length of the
latter. For finite plates,
the edge effect contributes to the Casimir force, effectively increasing the
plates' area \cite{Gies:2006xe, Weber:2009dp}.

The Casimir torque ${D_\mathrm{c}^\mathrm{i.p.,\varphi}}$ 
can easily be
obtained 
from
%
%
%
%
\begin{align} \label{T0-ip-torque-1}
{D_\mathrm{c}^\mathrm{i.p.,\varphi}}=
\frac{\mathrm{d}E_\mathrm{c}^\mathrm{i.p.,\varphi}}{\mathrm{d}\varphi}.
\end{align}
%
Fitting the numerical data for the torque to an odd polynomial in the vicinity
of $\varphi \to \pi/2 $, we obtain (see Fig. \ref{T0-ip-7})
\begin{align} \label{T0-ip-torque-2}
\frac{{D_\mathrm{c}^\mathrm{i.p.,\varphi\rightarrow \pi/2}} a^2}{L_y}\approx
0.00329\left(\frac{\pi}{2}-\varphi\right)+0.0038\left(\frac{\pi}{2}-\varphi\right)^3.
\end{align}
For the other limit $\varphi\!\rightarrow \! 0$, the Casimir torque diverges. The
leading order\cite{Weber:2009dp}, 
\begin{align} \label{T0-ip-torque-3}
  {D_\mathrm{c}^\mathrm{i.p.,\varphi\rightarrow 0}} \cong
\frac{L_y\pi^2}{2880 a^2\varphi^2}\approx0.00343 \frac{L_y}{a^2\varphi^2},
\end{align}
%
is an
excellent approximation to Eq.~(\ref{T0-ip-torque-1}) for $\varphi$ not too close
to $\pi/2$.
The divergent Casimir torque per length can be converted into a finite torque per
unit area which  leads
to the classical result for the torque
,
\begin{align} \label{T0-ip-torque-4}
  D_\mathrm{c}^\mathrm{\parallel,\varphi\rightarrow 0} =
\frac{A
L_{z}\pi^2}{960 a^4}\approx0.0103 \frac{A L_{z}}{a^4}.
\end{align}
Here, $A$ and $L_z$ denote the semi-infinite plate's area and the extent in $z$
direction, respectively.

A new characteristic contribution emerges from the edge effect
Eq.~(\ref{T0-ip-4}) which effectively changes the shape of the upper plate
seen by worldlines, as the upper plate appears to be higher near and at the edge itself. This leads to a contribution which
works against the standard torque~(\ref{T0-ip-torque-4}).
The correction to Eq.~(\ref{T0-ip-torque-4}) emerging from the edge effect
then is\cite{Weber:2009dp} $\approx -(0.003660 \pm 0.000038)L_{y}/a^2$.

\subsection{Inclined plates at finite temperature}
\label{sec:finite-temperature}

Decomposing the Casimir free energy at finite temperature $T=1/\beta$ into its
zero-temperature part $E_\mathrm{c}(0)$ and finite-temperature correction
$\Delta E_\mathrm{c}(T)$,
\begin{align}\label{FT-1}
E_\mathrm{c}(T)=E_\mathrm{c}(0)+\Delta E_\mathrm{c}(T),
\end{align}
is straightforward in the worldline picture by using the relation
(\ref{Int-3}). The finite-temperature correction is purely driven by the
worldlines with nonzero winding number, whereas the complicated
geometry-dependent part of the calculation remains the same for zero or finite
temperature.
The same statement holds for the Casimir force
$F_\mathrm{c}(T)=
F_\mathrm{c}(0)+\Delta F_\mathrm{c}(T)$.

In the following, we concentrate on the low-temperature limit, $aT\ll 1$. Full expressions for arbitrary temperature can be found in Ref.~\refcite{Weber:2009dp}. From dimensional analysis of Eq. (\ref{FT-1}), we would naively expect the Casimir energy to be of the form
\begin{align}\label{FT-1b}
E_\mathrm{c}(T)=E_\mathrm{c}(0)(1+c_1 aT+c_2 (aT)^2+c_3 (aT)^3+\dots).
\end{align}
No negative exponents should appear in (\ref{FT-1b}) since the thermal part of the energy has to disappear as $T\rightarrow 0$. Generically, the $T=0$ Casimir energy $E_{\mathrm{c}}(0)$ diverges for surfaces approaching contact $a\to0$. From \Eqref{FT-1b}, we would naively expect the same for the thermal correction. If, however,  sufficiently many of the first $c_i$'s in \Eqref{FT-1b} vanish, then the thermal part of the Casimir energy will be well behaved in $a$ without a divergence for $a\rightarrow 0$.

This is indeed the case for parallel plates ($c_1=c_2=0$, and $E_{\mathrm{c}}(0) \sim 1/a^3$) and for inclined plates ($c_1=0$, and $E_{\mathrm{c}}(0) \sim 1/a^2$). As a consequence, an extreme simplification arises: the thermal contribution in the low-temperature limit can be obtained by first taking the formal limit $a=0$ (only in the thermal contribution, of course).

In the following, we argue that there is no divergence in the thermal contribution in the limit $a\to0$ for general geometries:
Imagine a fancy geometry. The $a$-divergent part arises from the regions near the points (or lines or surfaces) of contact as $a\rightarrow 0$. The surfaces in these regions by construction bend away from each other. The thermal contribution can now be made larger by  flattening the surfaces in the contact region. Let~us  now substitute these regions by broader parallel plates. Then, the local thermal contribution to the Casimir energy of the original configuration will be clearly smaller than the finite thermal contribution of parallel plates. As~the latter does not lead to divergences for $a\to 0$, there can also be no divergence for the general curved case arising from the contact regions. Of course, infinite geometries may still experience an infinite thermal force, as it is the case for two infinitely extended parallel plates, but the local thermal contribution to the force density will be finite.

\begin{figure}[t]
\begin{center}
\includegraphics[width=0.6\linewidth]{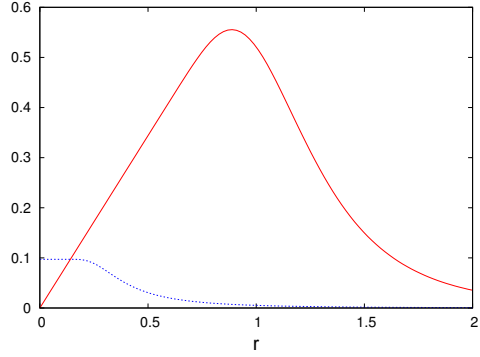}

\end{center}
\caption{The negative rescaled thermal force density Eq. (\ref{FT-1d})
  for perpendicular plates (dashed blue line) and the negative radial thermal force density for a sphere of radius $R=1$ above an infinite plate (solid red line) for $T=1$.
%
The thermal force density in the sphere-plate case has a maximum of $\approx 2\pi \times\pi^2/90$, where the factor $2\pi$ arises from the cylindrical measure. Note that a considerable fraction of the force density lies outside the sphere which only extends to $r=1$. As temperature drops, the maximum moves monotonously to the right.
}\label{T0-ip-FD}
\end{figure}

Another distinct feature of low-temperature effects is the spread of  the thermal force density  over regions of  size $\sim 1/T$ even for very small separations $a$. These effects are, of course, relevant for open geometries such as a sphere and cylinder above a plate\cite{our-Paper}. But it can also be demonstrated by calculating the thermal force density $\Delta f_\mathrm{c}(r,T)= f_\mathrm{c}(r,T)- f_\mathrm{c}(r,0)$ for two perpendicular plates at a distance $a=0$ as a function of the coordinate $r$ on the infinite surface measuring the distance from the edge (i.e., the contact point). The result can easily be obtained fully analytically on the worldline\cite{our-Paper}, yielding
\begin{align}\label{FT-1c}
\Delta f_\mathrm{c}(r,T)=-\frac{ T^4 L_y}{2 \pi^2}\!\left[\frac{\pi^4}{45}-\!\sum_{n=1}^\infty \! \left\langle\frac{2 e^{-(n \lambda_1/2 r  T)^2}}{n^4}-\frac{\lambda_1^2 e^{-(n\lambda_1/2 r T)^2}}{2 T^2 n^2 r^2}\right\rangle \right],
\end{align}
where $\lambda_1$ measures the extent of  half a unit worldline, i.e., the distance measured in $x$ direction from the left end to the center of mass. Rescaling~the radial coordinate  $r\rightarrow \lambda_1 r$ per worldline, the following rescaled force density leads to the same force upon integration over $r$,
\begin{align}\label{FT-1d}
\Delta \tilde{f}_\mathrm{c}(r,T)=-\frac{ T^4 L_y \sqrt{\pi}}{4 \pi^2}\left[\frac{\pi^4}{45}-\sum_{n=1}^\infty \left(\frac{2 e^{-(n /2 r  T)^2}}{n^4}-\frac{ e^{-(n/2 r T)^2}}{2 T^2 n^2 r^2}\right) \right],
\end{align}
where we have used $\langle\lambda_1\rangle=\sqrt{\pi}/2$. (Equations \eqref{FT-1c} and \eqref{FT-1d} possibly differ by a total derivative, but both provide for a reasonable thermal force density.) Upon integration, we obtain the thermal force of the perpendicular plates at $a=0$, $\Delta F_\mathrm{c}(T)=-\zeta(3)\,L_y T^3/4 \pi$ in agreement with Ref.~\refcite{Gies:2008zz}. In~the limit $a\to0$, the configuration has a scale invariance, which is reflected in the fact that Eqs.~(\ref{FT-1c}, \ref{FT-1d}) remain unchanged under $T\rightarrow  T\alpha , \ r\rightarrow r/\alpha $ and $\Delta \tilde{f}_\mathrm{c}\rightarrow \Delta \tilde{f}_\mathrm{c}/\alpha^4 $ for arbitrary $\alpha$. That means that evaluating (\ref{FT-1c}, \ref{FT-1d}) for say $T=1$ is sufficient to infer its form at all other $T$. Equation (\ref{FT-1d}) is shown for $T=1$ in Fig. \ref{T0-ip-FD}. The inflection point of each term in the $n$ sum is at $r_0=n/T\sqrt{10}$. For $r<1/T\sqrt{10}$ the force density stays nearly constant (and is equal to the first term in (\ref{FT-1c})) and rapidly goes  to zero for  $r> 1/T\sqrt{10}$. From this, we draw the important conclusion,  that the region of constant force density can be made arbitrarily large in $r$ direction by choosing  sufficiently low $T$.

Similar important consequences arise for temperature effects in other geometries. For example, the radial force density of a sphere above a plate exhibits a maximum due to the cylindric measure factor $r$, see Fig. \ref{T0-ip-FD}. Although this force density is not scale invariant due to the additional dimensionful scale $R$ (sphere radius), its maximum will nevertheless move away from the sphere as the temperature drops. No local approximate tools such as the PFA will be able to predict the correct thermal force. The fact that the force density is not scale invariant leads to  different temperature behaviors for $T<1/R$ and $T>1/R$ even in the limit $a\rightarrow 0$. \cite{our-Paper}

Let us now compare the low-temperature limit of the Casimir energy of two inclined plates with that of two parallel plates.
For $(aT)\rightarrow 0$, the correction $\Delta
E^{\parallel}_\mathrm{c}$ to the well-known parallel-plates energy reads
\begin{align}\label{int-sum-5}
\frac{\Delta E^{\parallel}_\mathrm{c}(aT\rightarrow 0)}{A}=-\frac{ \zeta(3)T^3}{4 \pi}+\frac{ \pi^2 a T^4 }{90}\approx-0.0957 T^3+0.110 a T^4.
\end{align}
Note that only the $T^4$ term
contributes to the force. 
The thermal correction 
to the inclined-plates energy is much more sensitive to temperature,
\begin{align}\label{int-sum-6}
  \frac{\Delta E_\mathrm{c}^{\mathrm{i.p.},\varphi}(aT\rightarrow 0)}{L_{y}}
  =-\frac{ \const_{\varphi,T_0} T^2}{24 \sin(\varphi)}+\frac{ \zeta(3) a T^3}{4 \pi \sin(\varphi)}, \end{align}
where $\const_{\varphi,T_0}$ was calculated numerically in Ref.~\refcite{Weber:2009dp} as a function of
$\varphi$. Only~the second term, which is a purely analytical result, contributes to the force. Equation (\ref{int-sum-6}) is
the generalization of a result for perpendicular plates,
  $\varphi=\pi/2$, see Ref.~\refcite{Gies:2008zz}.

Equation (\ref{int-sum-6}), being  an energy per edge length,
diverges as $\varphi\rightarrow 0$. As in \Eqref{T0-ip-4}, it has to be replaced by the energy of a semi-infinite plate above a parallel one,
$E_\mathrm{c}^\mathrm{1si}(T)
=E_\mathrm{c}^\mathrm{1si,edge}(T)+E_\mathrm{c}^\mathrm{1si,\parallel}(T)$. The
thermal part of $E_\mathrm{c}^\mathrm{1si,\parallel}(T)$ is as in
\Eqref{int-sum-5}, where $A$ is the area of the semi-infinite plate. The
leading thermal correction to the edge effect $\Delta
E_\mathrm{c}^\mathrm{1si,edge}(T)$ reads 
\begin{align}\label{int-sum-7}
\frac{\Delta E_\mathrm{c}^\mathrm{1si,edge}(T)}{L_y}=-\frac{ \const_{\varphi,T_0} T^2}{24}+ 0.063 a^{1.74}  T^{3.74}.
\end{align}
We find that the low-temperature regime of the 1si edge
effect is well described by a non-integer power law, $
\Delta F_{\text{c}}^{\text{1si,edge}}\sim T^{3.74}$,
where the fractional exponent arises from the geometry-temperature
interplay in this open geometry. Of course, our numerical analysis cannot
guarantee to
determine the true asymptotic behavior in the limit $a T\to 0$, but our data in the low-temperature domain $0.01\lesssim aT \lesssim 0.4$ are well fitted by the non-integer scaling also at higher fit orders\cite{Weber:2009dp}.  This result is very reminiscent to non-integer exponents known from critical phenomena. In both cases, this result arises from fluctuation contributions on all length scales, clearly revealing the long-range nature of both phenomena. Note that the leading temperature exponent of the 1si geometry is between the parallel-plates exponent $4$ and the inclined-plates exponent $3$, reflecting the fact that thermal properties of  the 1si geometry lie between those of the parallel and inclined plates.

The long-range nature of Casimir phenomena becomes also visible at the thermal correction to the torque. This is immediately transparent from Eq.~(\ref{int-sum-6}).
Whereas the $a$-independent first term of Eq.~(\ref{int-sum-6}) does not
contribute to the force, both terms in Eq.~(\ref{int-sum-6})
contribute to the low-temperature limit of the Casimir torque, the thermal
contribution being $
\mathrm{d}\Delta
  E_\mathrm{c}^\mathrm{i.p.,\varphi}(T)/\mathrm{d}\varphi$.
Concentrating on the limit $aT\rightarrow 0$ for small deviations from the perpendicular-plates case, $\varphi=\pi/2 {-} \delta \varphi$,
an expansion to first order in $\delta
\varphi$ yields:
\begin{align}\label{FT-ip9b}
\Delta  D_\mathrm{c}^{\mathrm{i.p.},\varphi=\pi/2 - \delta \varphi}(aT\rightarrow 0)/L_y
=\left(0.0716-0.0957 a T\right) T^2 \delta\varphi.
\end{align}
In the validity regime of the low-temperature expansion, $aT\ll
  1$, the positive first term is always dominant, hence the   perpendicular-plates
  case remains a repulsive fixed point. Most importantly, we would like to
  stress that the quadratic dependence of the torque on the temperature $\sim
  T^2$ ($\sim T^{D-2}$ in the general case) for the inclined-plates
  configuration represents the strongest temperature dependence of all
  observables discussed here.

\section{Conclusions}
\label{sec:conclusions}

A nontrivial interplay between finite temperature fluctuations and the geometry of a configuration can give rise to a variety of qualitatively different thermal corrections to Casimir phenomena. This effect becomes most pronounced in geometries where the relevant part of the spectrum is gapless. In these so-called open geometries, any small value of the temperature can excite low-lying thermal modes, giving rise to thermal corrections. By contrast, a gap in the relevant part of the spectrum of closed geometries suppresses thermal excitations at low temperature.

In the present work, we have developed the general picture underlying these geothermal Casimir phenomena. Open geometries, for instance, support a stronger influence of long-range fluctuations on thermal Casimir phenomena. We have illustrated the underlying mechanisms with the aid of the inclined-plates configuration and also presented first results for the experimentally important sphere-plate configuration.

Furthermore, we have presented a general argument that low-temperature corrections to Casimir forces become much more easily accessible by taking the (formal) contact limit $a\to 0$ (only for the thermal contributions), as thermal corrections remain well behaved in this limit. Whereas the existence of this limit is well known for parallel plates, we have argued that the same result holds for general geometries. The existence of this limit is also a reason why thermal corrections, for instance, in the perpendicular-plate case can be determined analytically. We expect that this observation will be useful for many other geometries as well. This should lead to practical simplifications also in other field theory approaches such as functional-integral approaches\cite{Bordag:1983zk,Emig:2001dx}, scattering theory\cite{Emig:2006uh,Kenneth:2006vr,Emig:2007cf,Rodrigues:2006ku,Milton:2007gy}, and mode summation\cite{Mazzitelli:2006ne}.

This particular geothermal interplay which we have observed in the context of the Casimir effect is certainly not restricted to Casimir physics. The crucial ingredients are a gapless fluctuation spectrum (though small gaps may not necessarily exert a strong quantitative influence) in a spatially inhomogeneous background. We expect that similar phenomena can occur for the thermal response of a system with an inhomogeneous condensate and an (almost) gapless fluctuation spectrum.

We conclude with the remark the geothermal interplay is only one out of several highly nontrivial interferences between deviations from the ideal Casimir limit.  For
instance, the interplay between dielectric material properties and finite
temperature \cite{Sernelius} is still {a subject of intense theoretical
  investigations} and has created a long-standing controversy
\cite{Mostepanenko:2005qh,Brevik:2006jw,Bimonte:2009nf,Ingold,Intravaia:2009pu}. Also the interplay between dielectric properties and geometry has been shown to lead to significant deviations from ideal curvature effects as well \cite{CanaguierDurand:2009zz}. All this exemplifies that a profound understanding of the Casimir effect requires a thorough quantum field theoretic basis.


\noindent {\bf Acknowledgments:} We would like to thank Kim Milton and all organizers of QFEXT09 for creating such a stimulating and productive conference atmosphere. This work was supported by the National Science.
 We~have benefited from activities within the
ESF Research Network CASIMIR and acknowledge support from the
Landesgraduiertenf\"orderung Baden-W\"urttemberg, the Heidelberg Graduate
School of Fundamental Physics (AW), and from the DFG grant Gi328/5-1 and
SFB-TR18 (HG).

\end{document}